\documentclass{article}
\usepackage{ijcai24}

\usepackage{times}
\usepackage{soul}
\usepackage{url}
\usepackage[hidelinks]{hyperref}
\usepackage[utf8]{inputenc}
\usepackage[small]{caption}
\usepackage{graphicx}
\usepackage{amsmath}
\usepackage{amsthm}
\usepackage{booktabs}
\usepackage{algorithm}
\usepackage{algorithmic}
\usepackage[switch]{lineno}
\usepackage{svg}
\usepackage{subcaption}
\usepackage{tabularx}
\usepackage{setspace}
\usepackage{multirow}
\usepackage{float}
\usepackage{tablefootnote}

\usepackage{color}

\title{AI-Olympics: Exploring the Generalization of Agents through Open Competitions}

\author{
Chen Wang$^{1}$
\and
Yan Song$^1$\and
Shuai Wu$^{2}$\and
Sa Wu$^1$\and
Ruizhi Zhang$^1$\and\\
Shu Lin$^1$\And
Haifeng Zhang$^{1,2,3,*}$\\
\affiliations
$^1$Institute of Automation, Chinese Academy of Sciences\\
$^2$Nanjing Artificial Intelligence Research of IA \\
$^3$School of Artificial Intelligence, University of Chinese Academy of Sciences\\
$^{*}$Corresponding Author\\
\emails
\{chen.wang, yan.song, sa.wu, ruizhi.zhang, chen.wang, shu.lin, haifeng.zhang\}@ia.ac.cn,\\
wushuai@airia.cn
}

\begin{document}

\maketitle

\begin{abstract}
    Between 2021 and 2023, \textit{AI-Olympics}---a series of online AI competitions---was hosted by the online evaluation platform Jidi in collaboration with the IJCAI committee. In these competitions, an agent is required to accomplish diverse sports tasks in a two-dimensional continuous world, while competing against an opponent. This paper provides a brief overview of the competition series and highlights notable findings. We aim to contribute insights to the field of multi-agent decision-making and explore the generalization of agents through engineering efforts.
\end{abstract}

\section{Introduction}

In recent years, AI-based decision-making has gained much attention in both academia and industries. Particularly, Reinforcement Learning (RL) has matured rapidly yielding high-level performance in various tasks in games \cite{Schrittwieser2019MasteringAG}, robotics \cite{Kober2013ReinforcementLI}, advertising \cite{Cai2017RealTimeBB}, etc. However, many of these carefully trained agents have been criticized for their poor generalization abilities when applied to a slightly difference task \cite{leibo2021scalable,gorsane2022towards,agapiou2022melting}. Implementing an adaptive agent that can perform equally well on a series of tasks is still an open and challenging problem. On the other hand,  assessing an agent's generalization skills normally relies on benchmark suits encompassing various environments  \cite{packer2018assessing,kirk2023survey}, tasks \cite{leibo2021scalable,oh2017zero} and opponents \cite{leibo2021scalable,agapiou2022melting}, which are often static and may prove inflexible amid the rapid evolution of agents.

To advance the research in this domain, we have developed a Python-based two-dimensional physical game engine \textit{AI-Olympics} environment, accompanied by a variety of scenarios. Meanwhile, we innovatively conducted a series of AI competitions utilizing diverse scenarios created within the framework for a more generalized evaluation. In this paper, we provide a brief introduction to \textit{AI-Olympics} environment and the accompanying scenarios, as well as the corresponding competition series and the evaluation of different generalization skills. A demonstration video is available at \url{https://youtu.be/SFXRe1JI6C8}.




\section{\textit{AI-Olympics} Environment}

\textit{AI-Olympics} environment is a two-dimensional physical simulator built from scratch with minimal dependencies, utilizing Pygame \cite{Shinde2021PygameDG} for visualization. The design ensures flexible deployment and extensibility. The underlying engine replicates the physical dynamics of agents and elastic collisions between geometric shapes, including straight lines, curves, and circles. Leveraging this game engine, various scenarios and tasks can be generated, spanning different configurations. Three essential characteristics define the environment: (1) \textbf{continuous control with partial observation}, (2) \textbf{zero-sum}, and (3) \textbf{multi-tasking}.

\subsection{Mobility and Observability}
\begin{figure}
    \centering
    \includegraphics[width=\linewidth]{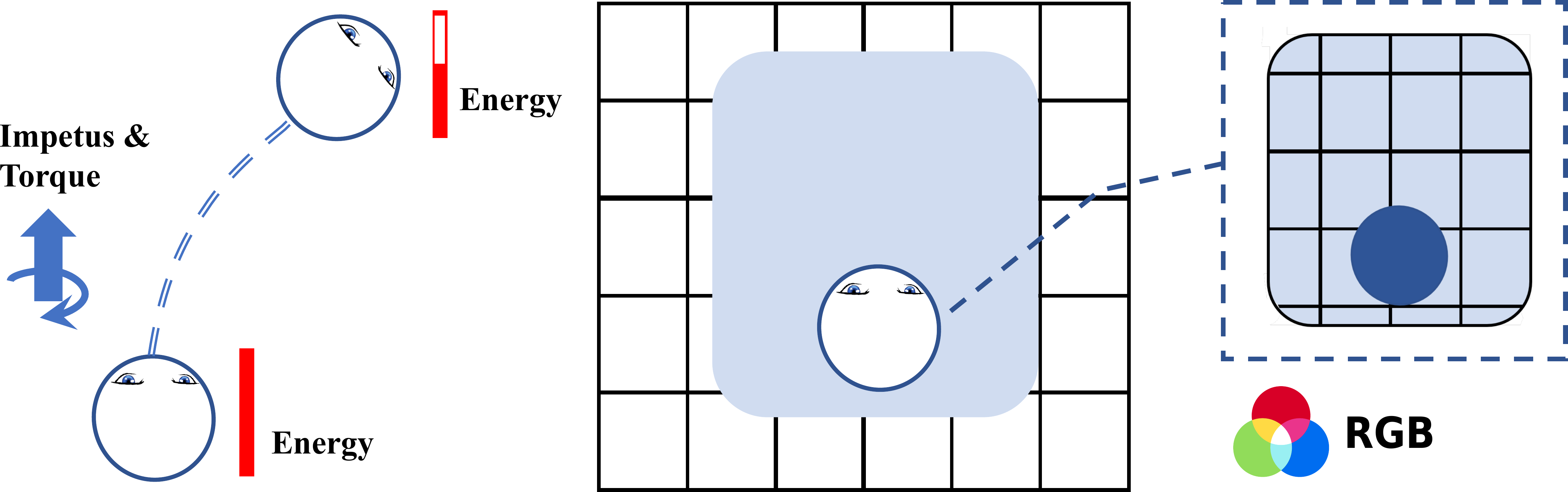}
    \caption{An illustration of agent's mobility and vision. An agent consumes stored energy and moves, while only observing partially of its surroundings. }\label{fig:agent_vision}
    \label{fig:enter-label}
\end{figure}

Each agent is depicted as an elastic circle on the map, equipped with the ability to exert impetus and torque, albeit at the cost of internal energy, enabling movement and interaction with the environment. Running out of energy results in a loss of control over the agent. Within the game environment, agents can collide and experience friction, thereby altering their states accordingly. As illustrated in Figure \ref{fig:agent_vision}, the agent can perceive its surroundings within a limited range, allowing it to observe any other object within the map. Other objects on the map are represented as geometric shapes, each distinguished by its unique color to denote specific characteristics. As a result, we can construct various maps with recognizable items placed, thereby enhancing intelligent AI recognition and generalization across scenarios.

\subsection{Zero-sum Games}

Typically, each game in \textit{AI-Olympics} contains two controllable agents with conflicting interests, rendering it a two-player zero-sum game extensively explored in prior research \cite{silver2016mastering,silver2017mastering,song2024empirical}. Both agents share the same mobility and observability for fair competition, and each can seek to gain an edge by obstructing pathways. Such uncertainty regarding the opponent's strategy during the evaluation phase can pose a significant challenge to the participants, encouraging them to consider generalization skills.


\subsection{Multiple Designed Tasks}\label{sec:scenarios}
\begin{figure}
    \centering
    \includegraphics[width=0.9\linewidth]{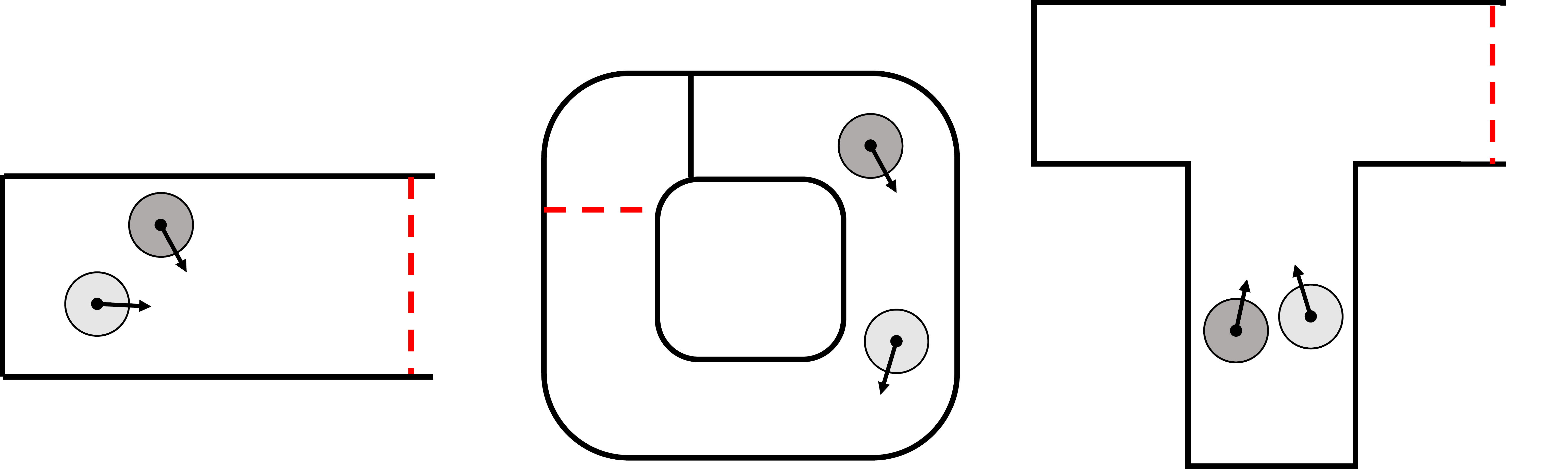}
    \includegraphics[width=0.7\linewidth]{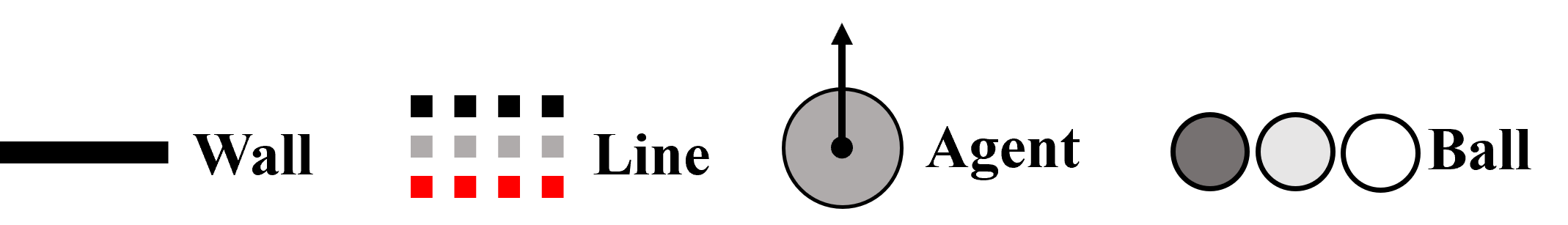}
    \caption{Three example maps in Running game. Agents are represented as circles and the arrow attached to them shows the driving force applied. }\label{fig:running}
\end{figure}

The game engine is crafted to generate diverse scenarios featuring common items on the map, allowing for an exploration of how agents can generalize their skills across tasks. We have developed six unique scenarios, each comprising a competitive sports task with different objectives and involving two agents with conflicting interests. These scenarios are:

\paragraph{Running game.} Running game involves two agents exploring a maze and seeking to reach the goal line (finish line) as fast as possible. Figure \ref{fig:running} gives three example maps with common map elements. For example, the black line represents a sticky wall where the agent loses speed when collisions, and the red dashed line represents the goal line. The first agent that reaches the goal line wins the game. 

\begin{figure}
    \centering
    \begin{subfigure}{0.3\linewidth}
        \centering
        \includegraphics[width=0.7\linewidth]{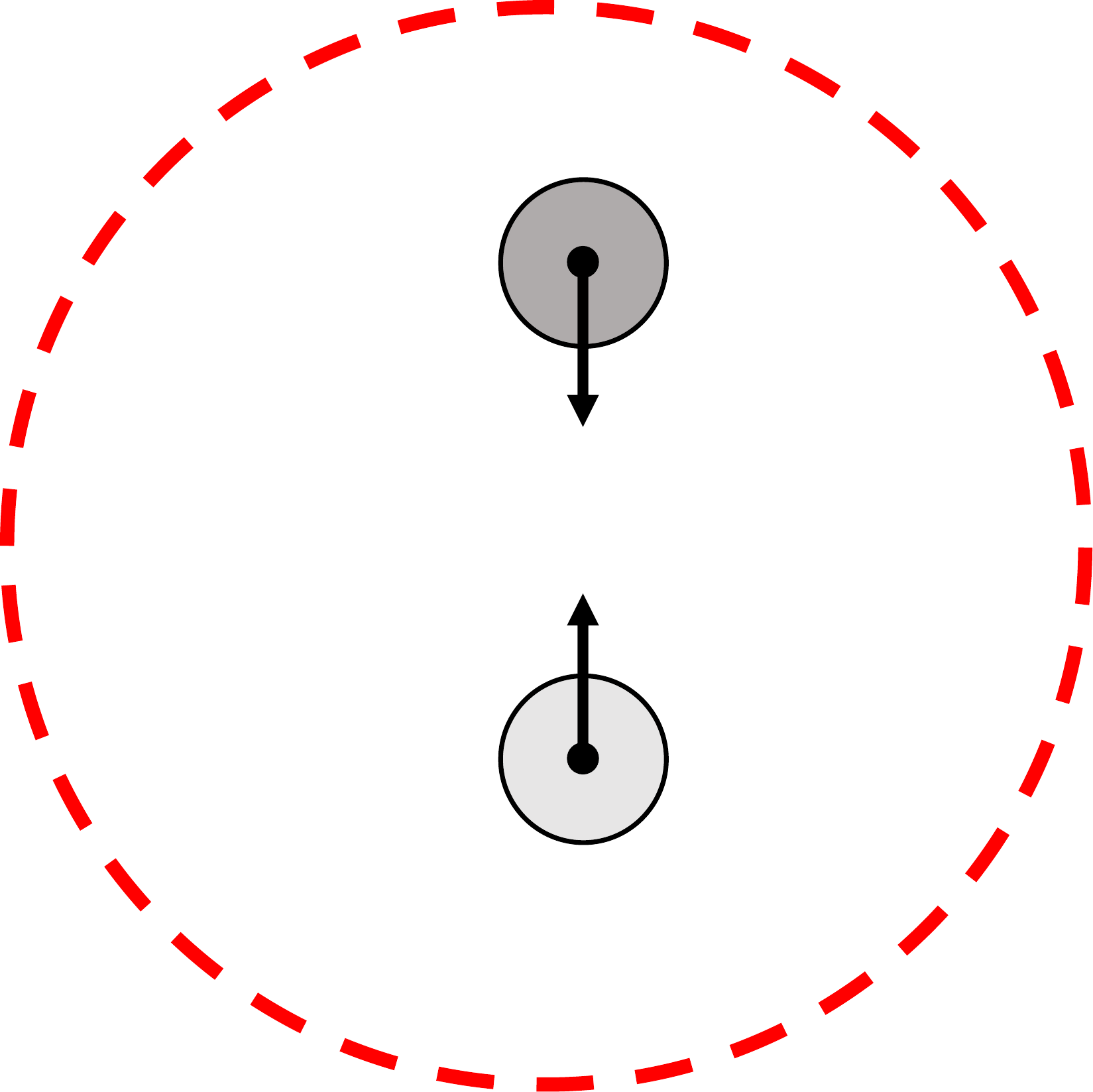}
        \caption{Wrestling}
    \end{subfigure}
    \begin{subfigure}{0.3\linewidth}
        \includegraphics[width=\linewidth]{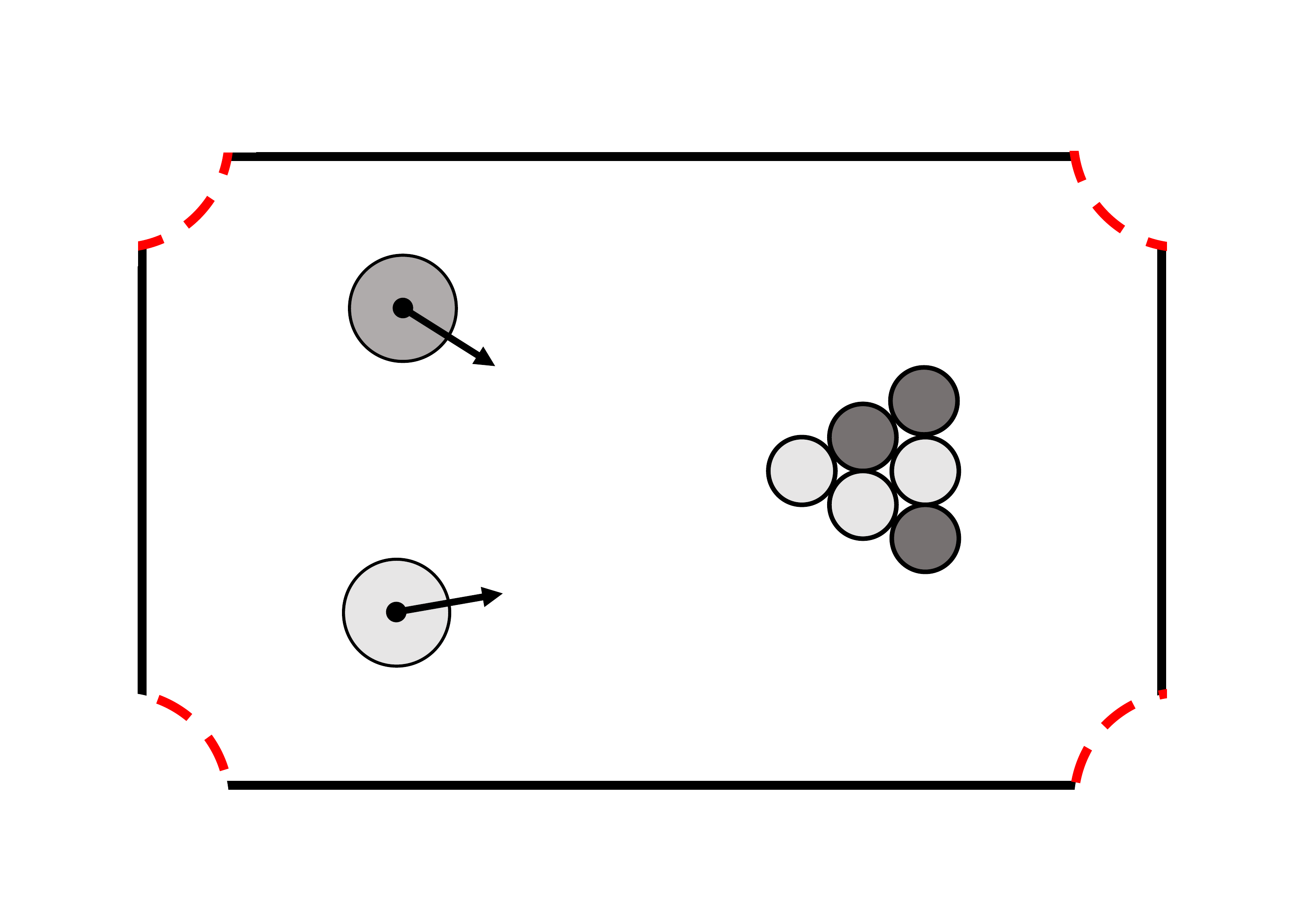}
        \caption{Billiard}
    \end{subfigure}
    \begin{subfigure}{0.3\linewidth}
        \raisebox{2.5mm}{
        \includegraphics[width=\linewidth]{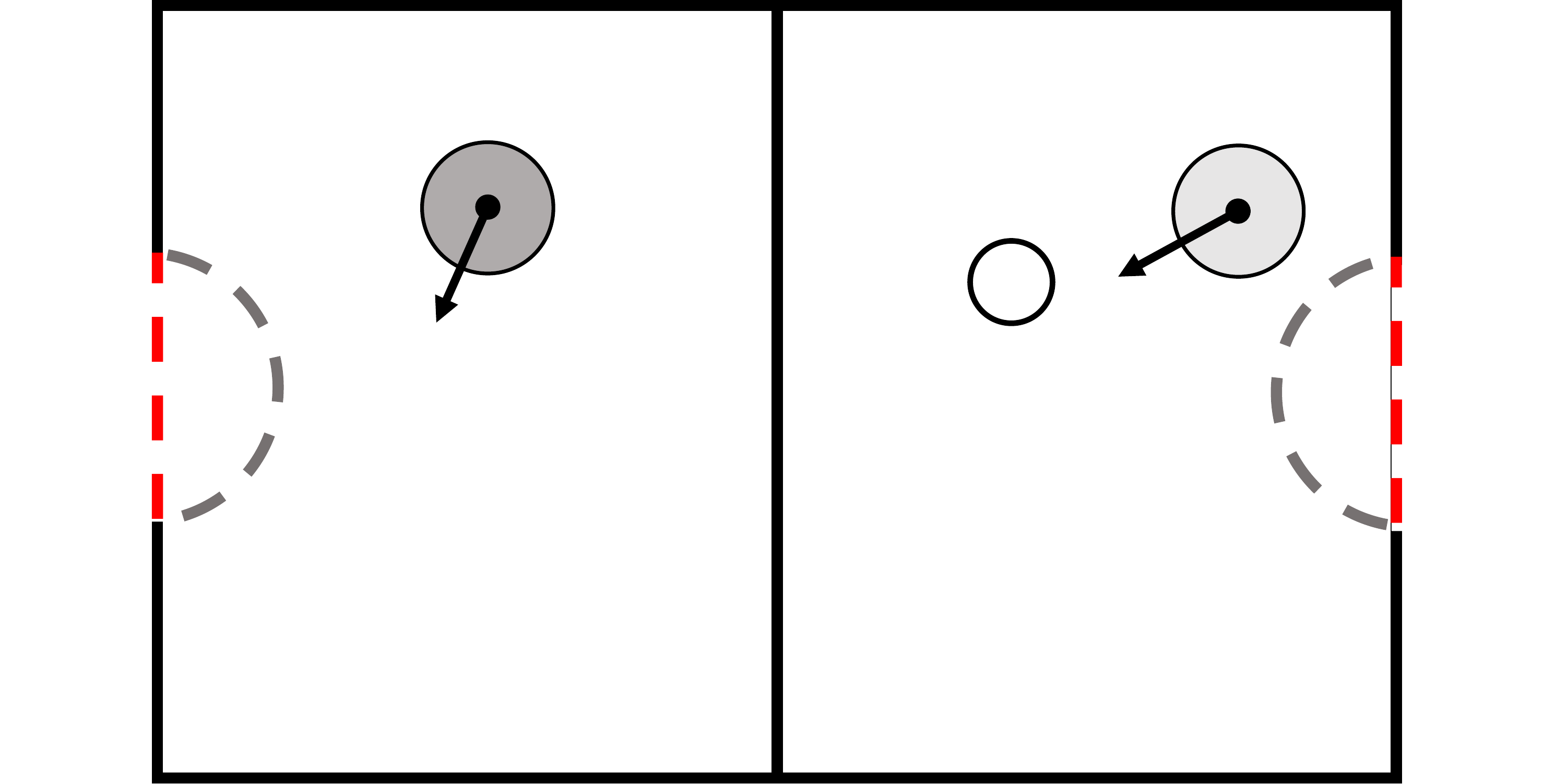}
        }
        \caption{Table-Hockey}
    \end{subfigure}
    \vskip 10pt
    \begin{subfigure}{0.3\linewidth}
        \centering
        \includegraphics[width=0.8\linewidth]{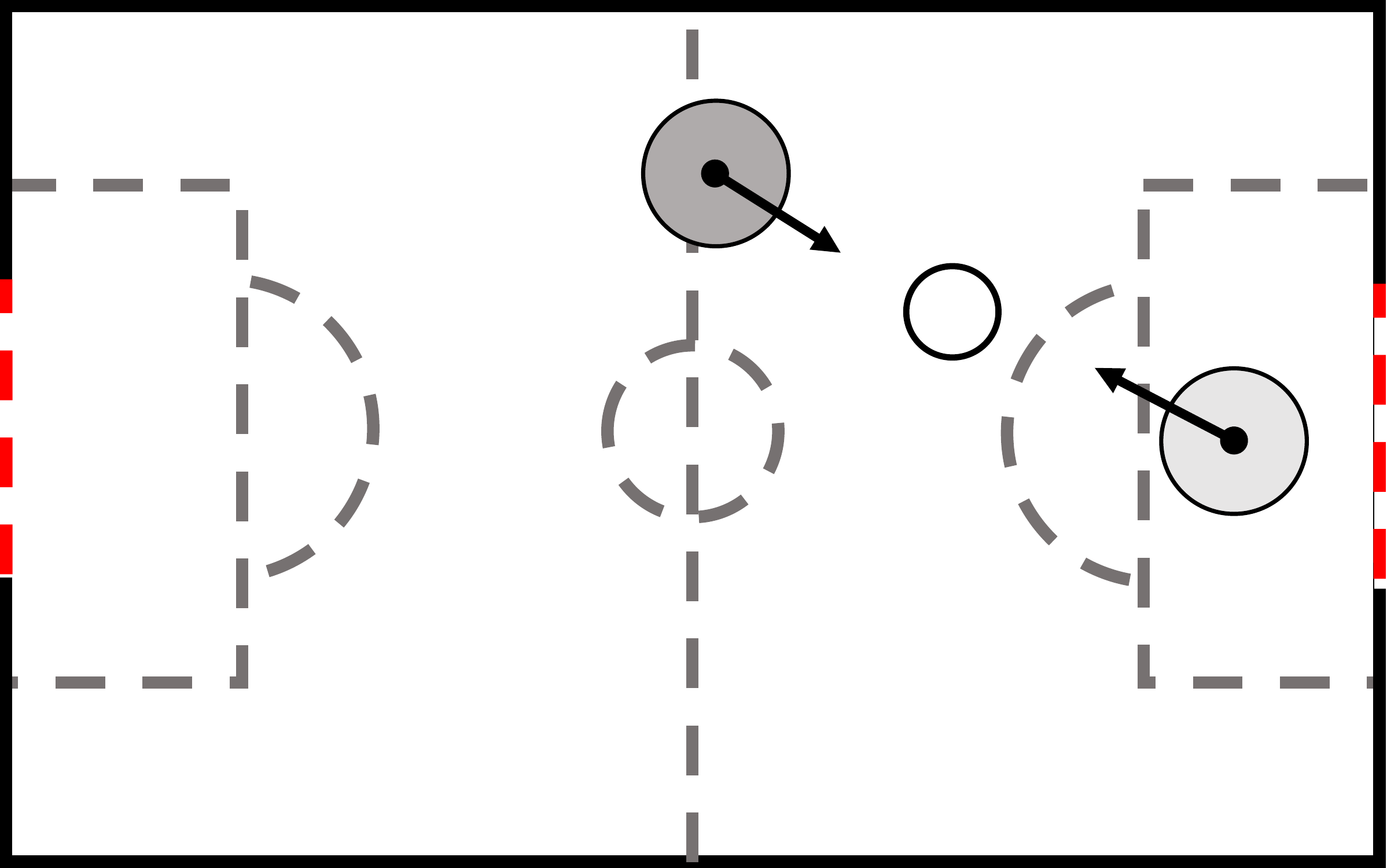}
        \caption{Football}
    \end{subfigure}
    \quad
    \begin{subfigure}{0.3\linewidth}
        \includegraphics[width=1.2\linewidth]{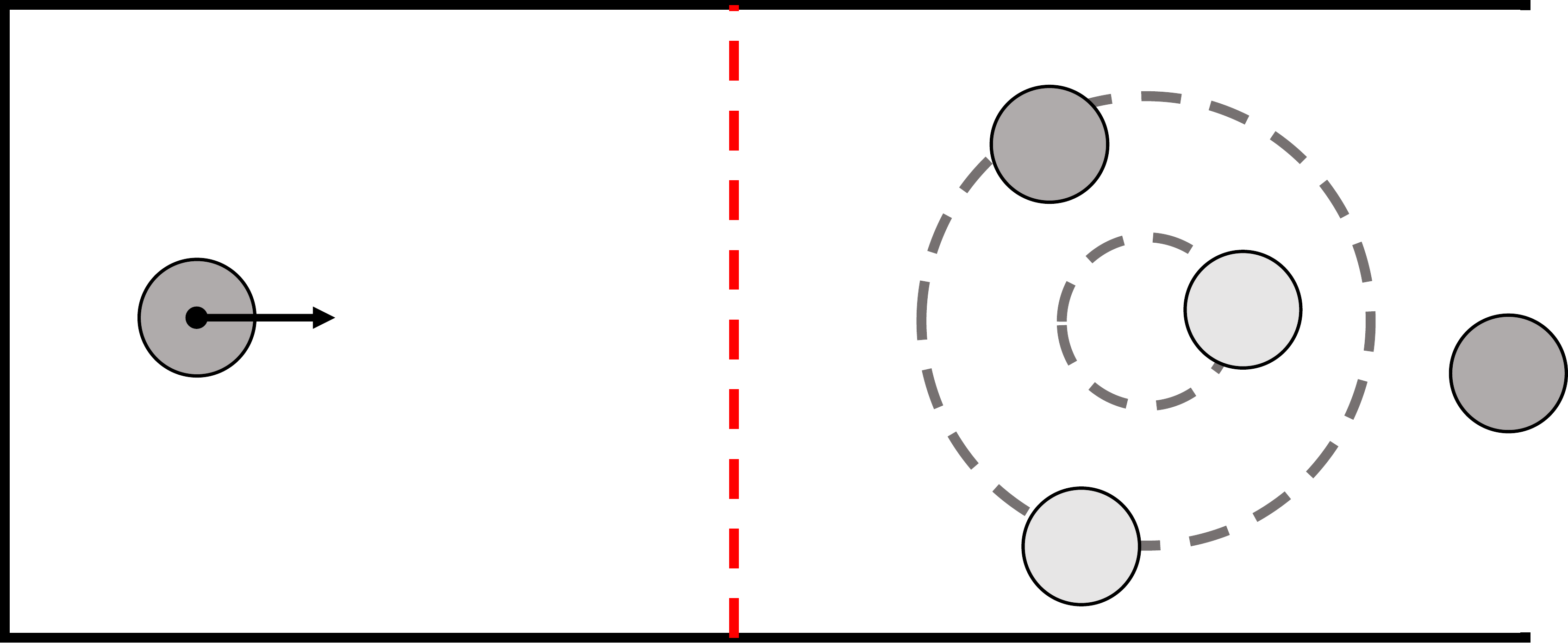}
        \caption{Curling}
    \end{subfigure}
    \includegraphics[width=0.7\linewidth]{fig/legend.png}
    \caption{Five additional scenarios built from \textit{AI-Olympics} game engine.}
    \label{fig:other tasks}
\end{figure}

\paragraph{Other single games.}  Five additional sports-related games are depicted in Figure \ref{fig:other tasks}. Wrestling game has the opposite objective compared to the Running game---each agent should keep itself away from the goal line (border line). In Curling game, players engage in a turn-based format where they strategically ``throw'' their rocks towards the central goal, aiming to position them as close to the center as possible. Both Table-Hockey and Football share similar map layouts, requiring players to score goals while also playing defensively. However, in Table-Hockey, agents are restricted to movement within their half of the field, while in Football, agents enjoy greater freedom of movement. In Billiard game, the objective of each agent is to pocket balls of the assigned color.

\paragraph{Integrated games.} The integrated game combines 4--6 distinct scenarios into one unified task. Two players engage in consecutive games, with the victor of the majority of games as the series winner. While each game features unique objectives, they are underpinned by shared elements including agent movement control and map items. Consequently, a well-trained agent can adeptly identify critical elements within its environment, enabling efficient exploration and navigation to overcome obstacles.


\section{\textit{AI-Olympics} Competition Series}

\textit{AI-Olympics} competition series adopts the tasks outlined in Section \ref{sec:scenarios} as themes and is evaluated by our online platform, \textbf{Jidi} (\url{http://www.jidiai.cn}). The platform offers live rankings and hosts AI competitions akin to Kaggle (\url{https://www.kaggle.com/}), specifically focusing on decision-making challenges. The objective of the competition series is to provide testbeds and promote practical engineering exploration of agents' generalization abilities. Notably, as shown in Table \ref{tab:jidi}, we have conducted several preliminary contests, two of which were partnered with the IJCAI committee to gain further attention.

\begin{table}
  \centering
    \begin{tabular}{ccc}
      \toprule
      \textbf{Contest}  & \textbf{Scenarios} & \textbf{Participants}  \\
      \midrule
        RLCN-2021-Winter     & Running          &    145 \\
        RLCN-2022-Spring          & Curling         & 103    \\
        RLCN-2022-Fall & Wrestling & 103   \\
        RLCN-2022-Winter & Table-Hockey  & 89  \\
        RLCN-2023-Spring & Billiard & 59  \\
        RLCN-2022-Summer  & Integrated(4)          & 166    \\
        \midrule
        \textbf{IJCAI-ECAI 2022} & Integrated(4) & 75 \\
        \textbf{IJCAI 2023} & Integrated(6) & 58 \\
      \bottomrule
    \end{tabular}
   \caption{Between 2021 and 2023, \textit{AI-Olympics} Competition Series has held eight online AI contests, among which six of them are preliminary and two are coordinated with the IJCAI committee. The competition series covers a different range of scenarios and has attracted numbers of participants. }\label{tab:jidi}
\end{table}


\subsection{Evaluation Protocol}

Platform Jidi offers online evaluation services for submitted agents in various simulated environments. The registered users submit their code files on the website, then the platform performs back-end evaluation and updates the results on the front-end webpage. Jidi is able to provide real-time ranking which is helpful for participants to test their agents. For \textit{AI-Olympics} competition series, since the games are all zero-sum, Jidi performs a \textit{Swiss-system Tournament} for player pairing during the evaluation process, which has been used in various types of competitions including sports tournaments and academic events \cite{csato2013ranking}.

\subsection{Generalization Abilities Assessment}
During the preliminary competition, we utilize individual scenarios as testbeds and invite participation in each specific task. Additionally, we explore an integrated scenario wherein the agent engages in all games sequentially. Throughout the competition, agents are assessed for their generalization abilities from various perspectives.

\paragraph{Map generalization.}
In the Running game competition, agents are required to efficiently navigate the track map and reach the goal line (finish line) before their opponent. The competition spans multiple rounds of evaluation, with newly designed maps added to the pool in each round. This ensures that the submitted agents are tested not only on familiar maps but also on unseen layouts. These new maps feature identical elements, as depicted in Figure \ref{fig:running}. A proficient agent must adeptly interpret the elements on the map and adjust its movement strategy accordingly, thereby demonstrating its generalization abilities across various maps.

\paragraph{Scenario generalization.} In the integrated games competitions, agents are required to compete in multiple sequential sports scenarios, each characterized by a distinct layout and objective described in Section \ref{sec:scenarios}. To prevent the agent from memorizing the scenario sequence, we employ random shuffling of orders, forcing the agent to identify the current scenario during the evaluation phase. Furthermore, a future extension may involve the addition of new tasks in each evaluation round, placing more emphasis on the agent's ability to generalize across scenarios.

\paragraph{Opponent generalization.} Throughout the competition series, the evaluation protocol requires the player not only solve multiple tasks but also compete with various opponents. Playing the same game with different opponents can yield disparate outcomes. Consequently, participants must take into account the strategies employed by other players. This can be achieved by reviewing replays of other players on the website and designing counter-strategies that can be generalized across various opponents.

\section{Analysis of the Submitted Agents}

During our competition series, some interesting findings echo our motivation as well as provide insights into the field.
    
    \paragraph{Enhancing map generalization through increased data.} In the Running game competition, at the beginning there are two maps in the candidate pool, then in subsequent rounds two additional maps are introduced into the pool, each with an increased probability of selection. As the competition progresses, we observe a consistent improvement in the mobility and completion rates of participants' agents on newly introduced maps. This phenomenon is attributed to participants incorporating the new maps into their training, thereby expanding the breadth of their training data through techniques such as data augmentation \cite{shorten2019survey}. Consequently, this broader scope of training data contributes to the enhancement of agents' generalization skills over time.
    
    \paragraph{Improving performance via scenario-specific strategies.} Despite the random order of scenarios, many agents still manage to apply a targeted policy to each scenario in the integrated game. In some scenarios such as Curling, which demands more intricate skills, participants demonstrated a notable trend for adapting their strategies to the current game situation. For instance, some agents will stop in front of the goal line (hog line) and observe the situation first and then move backward to leave space for ``throwing''. This aligns with our motivation to run the competition which is to ask the agent to act adaptively in the game instead of reacting to experience.

    \paragraph{Gaining advantages in competitive games through aggressive actions.} The presence of opponents introduces varying degrees of challenge. In scenarios emphasizing competition, we observe certain agents exhibiting interference behaviors. For example, in some maps of Running game, agents actively engage in tactics such as (1) intentionally colliding with opponents to slow their progress by pushing them against walls, or (2) strategically maneuvering to obstruct the paths of others during turns.

    \paragraph{Employing diverse AI-related methods.} After discussing with several participants, we have gained valuable insights into the methodologies they used to tackle the tasks at hand, including:
    \begin{itemize}
        \item reinforcement learning coupled with computer vision techniques,
        \item learning from demonstration to accelerate training,
        \item self-play framework for robustness,
        \item complex reward shaping to guide behavior,
        \item historical data summarization for long-term planning,
        \item curriculum learning to ease the training process, and
        \item heuristic methods to fill up the opponent pool.
    \end{itemize}
    In practice, participants often utilized a combination of these methods when implementing their agent. In contrast to state-of-the-art algorithms in academia, the competition underscored the practicality of machine learning methods and offered valuable engineering insights to the field of game AI.

\section{Conclusion}
\textit{AI-Olympics} competition series has provided a unique platform for participants to showcase their expertise in the field of game AI. Through the evaluation processes and diverse challenges, participants have demonstrated their proficiency in addressing complex scenarios and adapting to dynamic environments. The competition's emphasis on practicality and agent generalization abilities underscores the complexity of game AI research nowadays and brings insights for further research endeavors. In the future, we will keep exploring the generalization skills of agents and continue to gather community support to tackle the problem.

\section*{Acknowledgments}
Haifeng Zhang thanks the support of the National Natural Science
Foundation of China, Grant No. 62206289.




\bibliographystyle{named}

\end{document}